\documentclass[fleqn,twoside]{article}
\usepackage{espcrc2}
\input epsf 

\usepackage{graphicx}
\usepackage[figuresright]{rotating}

\newcommand{\AmS}{{\protect\the\textfont2
  A\kern-.1667em\lower.5ex\hbox{M}\kern-.125emS}}

\title{
       \vspace{-2.0cm}
       {\normalsize  ITEP-LAT/2002-24}     \\[-0.2cm]
       {\normalsize KANAZAWA 02-33}   \\[0.850cm]
The flux distribution of the three quark system in SU(3)} 

\author{H. Ichie\address{
        Institut f\"ur  Physik, Humboldt Universit\"at, 
        Invalidenstr.~110, D-10115 Berlin, Germany},
        V.~Bornyakov\address{ 
Institute for Theoretical Physics, Kanazawa University, Kanazawa
920-1192, Japan},
        T.~Streuer\address{
        NIC/DESY Zeuthen, Platanenallee 6, D-15738 Zeuthen, Germany}
        and 
        G.~Schierholz$^c$}
       
\begin{document}

\begin{abstract}
We study the abelian color-flux distribution of the three quark system
in the maximally abelian gauge on SU(3) lattices.
The distribution of the color electric field suggests $Y Ansatz$,
which might be interpreted through the dual superconductor picture
as the result of the vacuum pressure in the confined phase.
In order to clarify the flux structure,
we investigate the color electric field in the three quark system
also in the monopole part and in the photon part.
\vspace{1pc}
\end{abstract}

\maketitle

\section{Introduction}

The SU(3) lattice study on the three quark (3Q) system is
important for the clarification of the baryonic structure.
For more than 20 years, there is a crucial 
question about the long range force of the 3Q 
system \cite{sw,tes,bali,aft,afj,tmns}: 
whether there is a genuine three body force, 
or it can be described by the sum of the two body forces.
In the former case, the flux structure is expected to be of ``$Y$-shape'',
which has a junction at the point where the total length
of strings from a quark to the junction is minimal. 
On the other hand, in the latter case, the flux structure
is expected to be of ``$\Delta$-shape'', which consists of
the 3 sets of two body interactions.    

In order to distinguish between the two $Ans\ddot{a}tze$, the 3Q potential
was investigated using lattice simulations by several groups  
more than 15 years ago\cite{sw,tes} and also
recently\cite{bali,aft,afj,tmns}. 
The question, 
however, hasn't been settled yet. 
Some authors support the $\Delta Ansatz$\cite{bali,aft}, others find
evidence for the $Y Ansatz$ \cite{tmns} (see also \cite{afj}).
This discrepancy is due to the difficulty of measuring
the 3Q potential accurately for a large quark separation
and moreover the difference between the two $Ans\ddot{a}tze$ in the
3Q potential is not large, at most 20$\%$. 
Therefore, it is useful to measure the flux directly using
lattice simulations.

A couple of groups measured the flux in multi-quark systems.
However, for example, 
the Helsinki and the UK groups \cite{pgm} investigated them only in SU(2), 
where mesons and baryons have no differences, 
and no one has succeeded in the direct measurement of 
flux in the 3Q system in SU(3), since it needs huge statistics 
and large computer memory.
 
\begin{figure}[t]
\hspace*{1.5cm}
\hbox{
\epsfysize=5cm
\epsfxsize=5.cm
\epsfbox{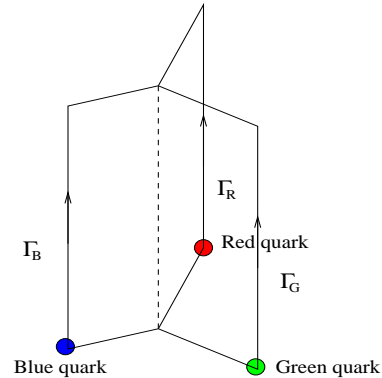}}
\vspace{-.5cm}
\caption{\it Three quark abelian Wilson loop.
}
\vspace*{-0.8cm}
\end{figure}

We shall study the flux of the three quark system in SU(3)
after fixing the gauge into the maximally abelian (MA) gauge
based on the dual superconductor scenario,
because  it is easy to interpret the flux structure
and the validity of this scenario for the flux tube has been confirmed by
Bali {\it et al.}\cite{bss}. In addition, one can get better signals 
on the abelian flux in the MA gauge than on the full flux in SU(3) 
and its computation needs only less statistics and memory than the full
one, which 
means that there is the possibility to succeed in the measurement of the flux
without waiting for a new bigger supercomputer.

\section{Operators and simulation details}

\begin{figure}[t]
\hspace*{0.5cm}
\hbox{
\epsfysize=8cm
\epsfxsize=5.cm
\epsfbox{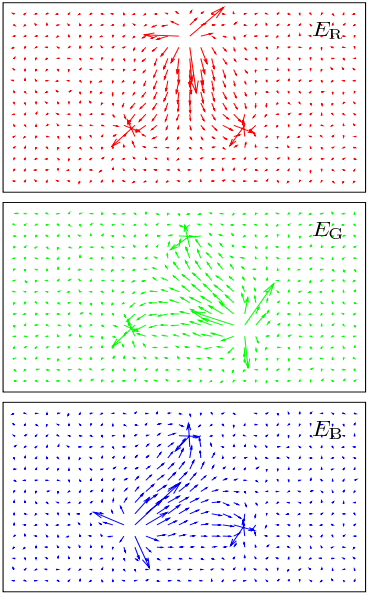}}
\vspace{-0.8cm}
\caption{\it The three components of the color electric field. 
The green component $E_{\rm G}$ flows from green quark
into red quark and blue quark.}
\vspace{-0.1cm}
\end{figure}

\begin{figure}[t]
\hspace*{-0.2cm}
\hbox{
\epsfysize=9.0cm
\epsfxsize=6.9cm
\epsfbox{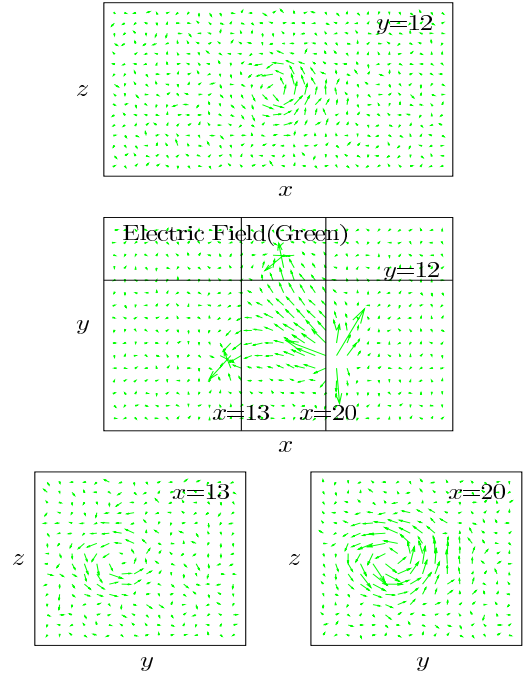}}
\vspace{-0.6cm}
\caption{\it The magnetic super currents rotating
around the color electric field (green component). }
\vspace{-0.1cm}
\end{figure}

The local quantities describing the abelian flux of the 3Q system
are obtained 
from the correlation between an appropriate operator and the 3Q abelian 
Wilson loop:
\begin{eqnarray}
O(s) = \frac{\langle O(s) W_{\rm 3Q} \rangle}{\langle W_{\rm 3Q} \rangle }
 - \langle O \rangle,  \nonumber
\end{eqnarray}
where the 3Q abelian Wilson loop is defined by 
\begin{eqnarray}
 W_{\rm 3Q} = \frac{1}{3!} |\epsilon_{abc}|
 u^a_{\rm R} \cdot u^b_{\rm G} \cdot u^c_{\rm B} \nonumber
\end{eqnarray}
with a path product 
\begin{eqnarray}
u^a_{ \mathcal C} = \prod_{s \in \Gamma_{\mathcal C} } u^a_\mu(s) \nonumber 
\hspace*{1cm}
\mbox{\large $\Gamma_{\mathcal C}= \Gamma_{\rm R},\Gamma_{\rm G},\Gamma_{\rm B}$}
\nonumber
\end{eqnarray}
of abelian link variables along the path $\Gamma_{\mathcal C}$ in fig. 1. 
Differently from the nonabelian case,
the color of quarks does not change during the propagation, 
because the off-diagonal components of the gauge field are frozen in the
abelian gauge.

The simulation is performed in quenched QCD at $\beta$=6.0 on
$16^3\cdot 32$ lattices. The link variables are fixed into the
maximally abelian gauge with a simulated annealing algorithm. 
The 3 static charges, red,
green and blue quarks, are located at $(17,14)$ $(22,6)$ and $(12,6)$
in the $x$-$y$ plane, respectively.
For noise reduction, we use the smearing method for spatial links of 
the 3Q abelian Wilson loop. 

\begin{figure}[t]
\hspace*{-0.7cm}
\hbox{
\epsfysize=9.0cm
\epsfxsize=7.7cm
\epsfbox{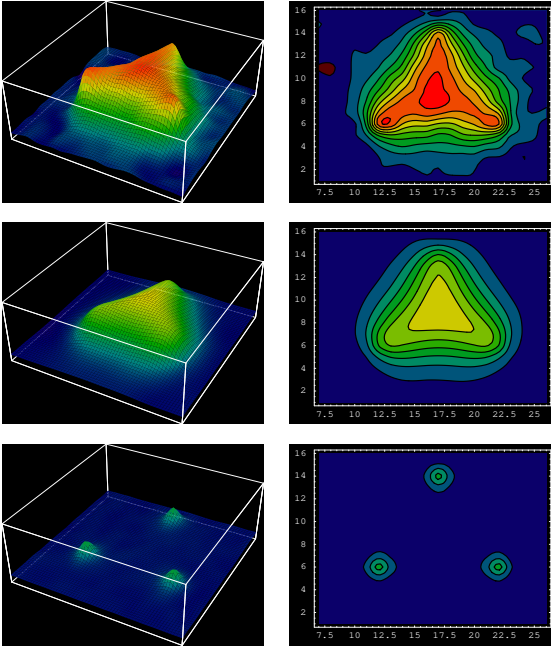}}
\vspace{-1cm}
\caption{\it Action density in the abelian(top),
monopole part(middle) and photon part(bottom).
}
\vspace{-0.6cm}
\end{figure}

\begin{figure}[t]
\vspace*{-0.9cm}
\hspace*{0.4cm}
\hbox{
\epsfysize=4.0cm
\epsfxsize=6.cm
\epsfbox{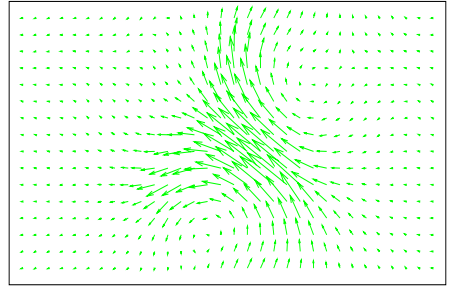}}
\vspace{-.5cm}
\caption{\it The induced electric field
in the monopole part (green component).  
}
\vspace{-0.6cm}
\end{figure}

\section{Flux of the 3 quark system}

Figs. 2 and 3 show the color electric field and the magnetic super current
in the 3Q system. 
For example, the green component
of the electric 
field flows from the green quark into 
the blue and red quarks.
For each flux, we observe
a solenoidal monopole current in the plane perpendicular to the 
flux. Corresponding to the amount of flux of the electric field,
there is a magnetic super current around each flux tube.   
Our results are in a qualitative agreement with dual Ginzburg-Landau
theory \cite{kist}.

In order to clarify the flux structure,
we decompose the abelian gauge field into the monopole part 
and photon part\cite{this}. 
Fig. 4 shows the action density in the abelian,
monopole and  photon parts, respectively.
The 3 peaks originating from the sources appear in 
the photon part, while the plateau, which is 
responsible for a confinement potential,  
appears in the monopole part. 
There does not appear a dent, which is expected
in the case of $\Delta Ansatz$, in the center of the
plateau. Then,  the abelian action density has a clear 
junction in the center of the 3 quarks, which suggests 
$Y Ansatz$. 

As shown in fig. 5, in the monopole part, there appear two rotating 
induced electric fields, which cancel the electric field
by original sources outside of the flux.  
Thus, we can expect the following scenario\cite{this}:
When the three quark sources are put into the vacuum, the
solenoidal monopole current appears for each flux, and
then they induce the two rotating electric fields such that
the flux is squeezed into the ``$Y$-shape''.

Moreover, we investigated the flux of the 3Q system 
also in the full QCD generated with the non-perturbatively 
$O(a)$ improved Wilson fermion action at 
$\beta$=5.29 and $\kappa$=0.1355\cite{booth}.  
For $m_\pi/m_\rho \simeq 0.7$, the flux structure is almost the same
and we did not observe any big effect of dynamical quarks, like
in the case of the quark-antiquark flux tube\cite{bikk}.  


$ $ 

\vspace{-0.1cm}
This work is partially supported by grant INTAS-00-00111.
H. I. thanks Humboldt University for hospitality. H. I. and V. B.
are supported by JSPS. Computations have been done on COMPAQ AlphaServer
ES40 at Humboldt University.








\end{document}